\documentclass[aps,prd,nofootinbib,twocolumn,superscriptaddress,preprintnumbers]{revtex4} % {revtex4-1}

\usepackage[dvipsnames]{xcolor}

\newif\ifshowtitles
\showtitlestrue
\def\ptitle#1{\ifshowtitles #1\fi}

\usepackage{amssymb}
\usepackage{amsmath}
\usepackage{epsfig}

\makeatletter
\def\simgt{\mathrel{\lower2.5pt\vbox{\lineskip=0pt\baselineskip=0pt
           \hbox{$>$}\hbox{$\sim$}}}}
\def\simlt{\mathrel{\lower2.5pt\vbox{\lineskip=0pt\baselineskip=0pt
           \hbox{$<$}\hbox{$\sim$}}}}
\makeatother

\newcommand{\be}{\begin{equation}}
\newcommand{\ee}{\end{equation}}
\newcommand{\bea}{\begin{eqnarray}}
\newcommand{\eea}{\end{eqnarray}}

\begin{document}

\preprint{MIT-CTP-4490, UCB-PTH-13/08}

\title{Inflationary Paradigm after Planck 2013}

\author{Alan H. Guth}
\affiliation{Center for Theoretical Physics, Laboratory for Nuclear Science,
 and Department of Physics, Massachusetts Institute of Technology, 
 Cambridge, MA 02139, USA}

\author{David I. Kaiser}
\affiliation{Center for Theoretical Physics, Laboratory for Nuclear Science,
 and Department of Physics, Massachusetts Institute of Technology, 
 Cambridge, MA 02139, USA}

\author{Yasunori Nomura}
\affiliation{Berkeley Center for Theoretical Physics, Department of Physics,
 and Theoretical Physics Group, Lawrence Berkeley National Laboratory,
 University of California, Berkeley, CA 94720, USA}

\begin{abstract}
Models of cosmic inflation posit an early phase of accelerated expansion 
of the universe, driven by the dynamics of one or more scalar fields in 
curved spacetime.  Though detailed assumptions about fields and couplings 
vary across models, inflation makes specific, quantitative predictions 
for several observable quantities, such as the flatness parameter ($\Omega_k 
= 1 - \Omega$) and the spectral tilt of primordial curvature perturbations 
($n_s - 1 = d \ln {\cal P}_{\cal R} / d \ln k$), among others---predictions 
that match the latest observations from the {\it Planck} satellite to very 
good precision.  In the light of data from {\it Planck} as well as recent 
theoretical developments in the study of eternal inflation and the multiverse, 
we address recent criticisms of inflation by Ijjas, Steinhardt, and Loeb. 
We argue that their conclusions rest on several problematic assumptions, 
and we conclude that cosmic inflation is on a stronger footing than 
ever before.
\end{abstract}

\date {December 29, 2013, revised January 13, 2014}

\maketitle

\section{Introduction}
\label{sec:intro}

Did our universe undergo a period of accelerated expansion in the 
early stage of its evolution?  If so, does it play an important role 
in explaining observable features of our universe today?

We define the ``inflationary paradigm'' to mean that the answer to 
both of these questions is ``yes''~\cite{Guth:1980zm,Linde:1981mu}. 
As we argue here, the inflationary paradigm draws upon well-motivated 
physical interactions and types of matter.  The inflationary explanations 
for the homogeneity and the flatness of the universe can be understood 
in the context of classical general relativity, and even the 
origin of density fluctuations can be accurately described 
in the context of quantum field theory on a classical, curved 
spacetime~\cite{Guth-Solvay}, a theoretical framework that has 
been thoroughly studied for decades~\cite{books}.  Moreover, reasoning 
about the behavior of fundamental scalar fields is on a stronger 
footing than ever, in the light of the recent observation of the 
Higgs boson at the LHC~\cite{Aad:2012tfa,PDGupdate2014}.

As is well known, inflation makes several generic 
predictions~\cite{Guth:2005zr,Tegmark:2004qd}.  The observable 
universe today should be flat, i.e., $|\Omega_k| \ll 1$, where 
$\Omega_k \equiv 1 - \Omega$.  There should exist primordial 
curvature perturbations whose power spectrum ${\cal P}_{\cal R}(k) 
\sim k^{n_s-1}$ has a slightly tilted spectral index, $|n_s-1| \ll 1$, 
typically red-shifted.  Unless the inflaton potential or the initial 
conditions are fine-tuned, the primordial perturbations should be 
predominantly Gaussian~\cite{Bartolo:2004if}.  Modes of a given (comoving) 
wavelength should ``freeze out'' upon first crossing the Hubble radius 
during inflation, remain (nearly) constant in amplitude while longer 
than the Hubble radius, and then resume oscillation upon reentering 
the Hubble radius.  The temporal oscillations of modes with nearby 
wavelengths are therefore coherent~\cite{Dodelson:2003ip}, giving 
rise to a sharp pattern of peaks and troughs in the cosmic microwave 
background (CMB) power spectrum.  These generic predictions are 
consequences of simple inflationary models, and depend only on the 
physics at the inflationary energy scale, i.e., the energy scale of 
the final stage of inflation, as observed in the CMB\@.  We will refer 
to these as inflation-scale predictions.  To date, every single one of 
these inflation-scale predictions has been confirmed to good precision, 
most recently with the {\it Planck} satellite~\cite{Ade:2013uln}.

Despite these successes, Ijjas, Steinhardt, and Loeb 
(ISL)~\cite{Ijjas:2013vea} have recently argued that the inflationary 
paradigm is in trouble in the light of data from {\it Planck}. 
They agree that a class of inflationary models make predictions 
that agree with experiment, which is how theories are usually evaluated, 
but they bring up a different issue.  They argue that if one starts at 
the Planck scale with reasonable assumptions about initial conditions, 
the successful inflationary models are ``exponentially unlikely 
according to the inner logic of the inflationary paradigm itself.'' 
In this paper we argue that this is not the case by addressing each 
of their specific points.  We will argue that their negative conclusions 
rely on unfounded assumptions, and can be completely avoided under 
what we consider to be more reasonable assumptions about the physics 
between the inflationary scale and the Planck scale.

We also believe, as a matter of principle, that it is totally 
inappropriate to judge inflation on how well it fits with anybody's 
speculative ideas about Planck-scale physics---physics that is well 
beyond what is observationally tested.  All theories of evolution begin 
with assumptions that are taken to be plausible, but which are usually 
not directly verifiable, and then the theories make predictions which 
can be tested against current observations.  We do not reject Darwinian 
evolution because it does not explain the actual origin of life; we 
do not reject big-bang nucleosynthesis because it does not explain the 
homogeneous thermal equilibrium initial state that it requires; and we 
should similarly not even consider rejecting the inflationary paradigm 
because it is not yet part of a complete solution to the ultimate 
mystery of the origin of the universe.  For us, the implications go 
the other way:\ the successes that inflation has had in explaining 
the observed features of the universe give us motivation to explore 
the speculative ideas about the implications of inflation for questions 
far beyond what we can observe.

If inflation occurred in the early universe, then the evidence of its 
own initial conditions would be effectively erased, as described by the 
cosmic no-hair conjecture~\cite{Wald:1983ky}.  Thus, the earliest moments 
of inflation, or anything that might have come before, are extremely 
difficult to probe observationally.  Nonetheless, the inflationary 
framework does provide resources with which to address important open 
questions, such as the initial conditions at or near the Planck scale. 
Within that framework, important advances have been made in recent years 
on topics such as eternal inflation~\cite{Guth:1982pn}, the multiverse 
and various proposals to define probabilities~\cite{Bousso:2000xa,%
DeSimone:2008bq,Linde:1993xx,Bousso:2006ev,Nomura:2011dt,Garriga:2012bc}, 
and the possible role of anthropic selection effects~\cite{Weinberg:1987dv,%
Barrow-Tipler,Agrawal:1997gf}.  Most important, as we discuss below, 
the inflationary paradigm has expanded beyond what was once the dominant 
view, prevalent in the 1980s, which tended to focus on a single phase 
of ``chaotic'' inflation~\cite{Linde:1983gd}.  Given recent progress 
on both the observational and theoretical fronts, we believe that the 
inflationary paradigm is in far better shape than ever before.

The remainder of this paper is organized as follows.  In 
Section~\ref{sec:Planck}, we discuss the implications of the Planck~2013 
data.  In Section~\ref{sec:initial} we discuss the initial conditions 
for inflation, in Section~\ref{sec:multiverse} we discuss the issue 
of predictions in the multiverse, in Section~\ref{sec:unlikeliness} 
we discuss what ISL call the inflationary ``unlikeliness problem,'' 
and in Section~\ref{sec:LHC} we discuss the possibility that the 
Higgs potential turns negative at large field values.  We summarize 
in Section~\ref{sec:concl}.

\section{Planck 2013 Data}
\label{sec:Planck}

ISL argue that the Planck~2013 data prefers single-field inflation over 
more complicated possibilities, and that a ``plateau-like'' potential looks 
better than other simple potentials such as power-law potentials.  They 
argue that these facts lead to significant challenges to inflation.

The relevant observational constraints on the shape of the potential come 
from $r$, the ratio of the power spectra of tensor and scalar perturbations. 
For single-field models, $r$ is proportional to the slow-roll parameter 
$\epsilon \equiv - \dot{H} / H^2$ and hence to $(V_{,\phi} / V )^2$.  Thus 
small values of $r$ require modest slope of the inflationary potential, 
at least in the vicinity of $\phi_I \equiv \phi(t_I)$, where $t_I$ is 
the time during inflation when cosmologically relevant length scales 
first crossed outside the Hubble radius.

On their own, the {\it Planck} data constrain $r < 0.12$ at the pivot 
scale $k_* = 0.002~{\rm Mpc}^{-1}$ at $95\%$~CL~\cite{Ade:2013uln}. 
This bound represents an impressive improvement from the WMAP9 constraint 
($r < 0.38$~\cite{Hinshaw:2012aka}), although it is comparable to the 
constraints that arise from combining WMAP data with data from the 
South Pole Telescope (SPT) and measurements of the baryon acoustic 
oscillations (BAO): $r < 0.18$ for WMAP7+SPT and $r < 0.11$ for 
WMAP7+SPT+BAO~\cite{Story:2012wx}.  The {\it Planck} constraint is 
little changed if one incorporates data from SPT, BAO, the Atacama 
Cosmology Telescope (ACT), and large-scale polarization data from 
WMAP9; these combinations yield $r < 0.11 - 0.13$~\cite{Ade:2013uln}.

The constraint $r < 0.12$ is low enough that the simple, single-field 
model with $V = \lambda \phi^4$ falls outside the $95\%$~CL contour 
if one makes the usual assumptions about reheating and the thermalization 
energy scale after inflation.  Another simple model, with $V = 
\frac{1}{2} m^2 \phi^2$, lies at the boundary of the $95\%$~CL 
contour, although it moves more squarely into the allowed region 
if the pivot scale corresponds to $N_* = 63$ $e$-folds before the 
end of inflation~\cite{Dodelson:2003vq} rather than $N_* \leq 60$.

Thus the latest data, while certainly impressive, hardly rule out simple 
models with polynomial potentials, although they do constrain parameter 
space at the $1\sigma$~--~$2\sigma$ level.  Nonetheless, ISL raise the 
conceptual question of whether plateau-like potentials are evidence 
against the inflationary paradigm.  The main point of this paper is 
to argue that {\it even if} the final stage of inflation, as observed 
in the CMB, is determined definitively to occur on a plateau-like 
potential, the inflationary paradigm is not in trouble at all.  As 
we discuss in the next section, the preferred scenarios might simply 
depart from a view about the onset of inflation that was commonly 
held two to three decades ago.

\section{Initial Conditions}
\label{sec:initial}

In this section we will assume, for the purpose of discussing ISL's 
conclusions, that the observable phase of inflation---the phase which 
we believe produced the density perturbations that we now measure in the 
CMB---indeed occurred on a ``plateau-like'' potential.  The constraints 
on $r$ discussed above then require the height of the plateau $V_I$ 
to be no bigger than about $10^{-12} M_{\rm Pl}^4$, where $M_{\rm Pl} 
\simeq 1.22 \times 10^{19}~{\rm GeV}$ is the Planck scale.  Because 
this energy density is so low, ISL argue that one needs very fine-tuned 
initial conditions at the Planck scale in order to have an approximately 
homogeneous region of Hubble size after the energy density falls to 
the needed value.  In particular, they argue that one cannot use the 
simple chaotic picture $\frac{1}{2} \dot{\phi}^2 \sim \frac{1}{2} 
|\nabla \phi|^2 \sim V$ near $\phi \sim M_{\rm Pl}$ to start the 
observable inflation, since the plateau potential energy density 
cannot be that high.  With $\frac{1}{2} \dot{\phi}^2 \sim \frac{1}{2} 
|\nabla \phi|^2 \gg V \sim 10^{-12} M_{\rm Pl}^4$ at the Planck era, 
ISL argue that a Hubble-sized region of homogeneity at the onset of 
inflation would require a region of homogeneity at the Planck scale 
of at least 1000 Hubble lengths.

We do not agree with this estimate, which in our view is based on 
false assumptions.  A very plausible way to cool from the Planck 
scale to energy densities of order $V_I$, while maintaining homogeneity, 
is to imagine starting from a region of negative spatial curvature, 
$k < 0$, so that it locally resembles an open Friedmann-Robertson-Walker 
universe.%
\footnote{We thank Alex Vilenkin for pointing this out.
Alternatively, universes with nontrivial topology, such as a
torus, can also cool from the Planck scale to low energies while
maintaining homogeneity~\cite{Linde:2004nz}.  In this scenario it is
even possible for initial inhomogeneities to be smoothed by
``chaotic mixing''~\cite{Linde:2004nz,Cornish:1996st}.}
Note that $k=0$ would be a very special case, and that regions with 
$k>0$ would recollapse before reaching $V_I$, unless they were very 
close to being flat.  The curvature term in the Friedmann equation, 
like the gradient energy $\frac{1}{2} |\nabla \phi|^2$, scales as 
$1/a^2(t)$, where $a(t)$ is the scale factor.  The scalar field kinetic 
energy $\frac{1}{2} \dot{\phi}^2$ scales as $1/a^6(t)$, so the $1/a^2(t)$ 
terms will quickly dominate, leading to the behavior $a(t) \propto t$. 
If we make the plausible assumption that a region of homogeneity grows 
with the expansion of the universe, then both the size of the region 
and the Hubble length grow proportionally to $t$, and a Hubble-sized 
region at the onset of inflation requires no more than a Hubble-sized 
region at the Planck scale.

While the above argument seems reasonable, critics might argue 
that we are being overly optimistic, because perhaps the comoving 
size of the region of homogeneity might shrink as the universe 
evolves.  The worst case would be a scenario in which the inhomogeneity 
from outside the region propagates inward, limited only by the speed 
of light.%
\footnote{The claim that this is the worst possible case can be 
 investigated rigorously in the context of classical general relativity, 
 and the relevant theorems are discussed in detail by Wald~\cite{WaldBook}. 
 For Einstein's equations in vacuum, Theorem~10.2.2 implies as a special 
 case that if an initial spacelike slice contains a region $S$ described 
 by a Robertson-Walker metric, then the future domain of dependence 
 of $S$---the region calculated in Eqs.~(\ref{radius1}) and 
 (\ref{radius2})---will be unperturbed, regardless of what is 
 outside the region $S$.  Conditions outside of $S$ cannot affect 
 anything inside the future domain of dependence, and cannot affect 
 where the boundary of that region occurs.  On pp.~266 and 267, Wald 
 discusses extensions of this theorem when matter is included, at 
 least for simple forms of matter.  While the theorem has not been 
 proven for all forms of matter, we think it is safe to assume that 
 any {\it acceptable} theory of matter would satisfy these basic 
 causality properties.}
In that case, the physical radius $r(t_I)$ of the region of homogeneity 
at the onset of inflation can be related to the radius $r(t_{\rm Pl})$ 
of homogeneity at the Planck scale by
\begin{equation}
  r(t_I) = a(t_I) \left[ {r (t_{\rm Pl}) \over a(t_{\rm Pl})} 
    - \int_{t_{\rm Pl}}^{t_I} {d t \over a(t)} \right] \ . 
\label{radius1}
\end{equation}
If we set $r(t_I) = H^{-1}(t_I)$ and $H^2(t_{\rm Pl}) \equiv {8 \pi \over 3} 
M_{\rm Pl}^2$, and assume that $a(t) \propto t$, the above equation becomes
\begin{equation}
  r(t_{\rm Pl}) = {1 \over H(t_{\rm Pl})} \left[ 1 + 
    \ln \left( {H(t_{\rm Pl}) \over H(t_I)} \right) \right] 
  > 13.9 H^{-1}(t_{\rm Pl})\ ,
\label{radius2}
\end{equation}
where we have used the Planck~2013 95\% CL constraint~\cite{Ade:2013uln} 
that $H(t_I) < 3.7 \times 10^{-5} M_{\rm Pl} / \sqrt{8\pi}$.  Thus, even 
in this worst case scenario, the factor of 1000 given by ISL is replaced 
by a factor of 13.9.  Synthesizing the calculations described in this 
paragraph and the previous one, we conclude that a Hubble-sized homogeneous 
region at the onset of inflation requires only a region of homogeneity 
at the Planck scale of order 1--15 Hubble lengths.

Besides our disagreement about the required size of the region of 
homogeneity at the Planck scale, we more significantly disagree with 
the entire premise of the argument.  ISL's argument is predicated on the 
assumption that the final stage of inflation---whose last $N \sim 60$ 
$e$-folds correspond to the observable inflation---was the end of an 
uninterrupted phase of inflation that began at the Planck scale.  That 
requirement is tantamount to assuming that $V (\phi)$ is essentially 
featureless between the values of $\phi$ at the Planck era and the 
era of observable inflation.  Given recent developments in high-energy 
theory (e.g.\ the revised understanding of the vacuum structure in 
string theory~\cite{Bousso:2000xa} and the idea that the effective 
theory below the Planck scale may contain multiple---often 
separate---sectors~\cite{Dimopoulos:2001ui}), we find it very 
plausible that $V(\phi)$ is much more complicated than that, with 
multiple fields and many local minima.  Thus we see little reason 
to expect (let alone require) that a single phase of early-universe 
inflation stretched all the way from the Planck to the observable 
inflation eras.

For example, the final stage of inflation could plausibly have begun 
by tunneling from some other metastable state.  In that case, the 
inflation in the previous metastable state together with the symmetry 
of the Coleman--De~Luccia instanton~\cite{Coleman:1980aw} would ensure 
spatial homogeneity (small $ |\nabla \phi|^2$) {\it prior to} the last 
stage of inflation.  (Since the bubble nucleation rate is exponentially 
suppressed, it is highly likely that the field before the tunneling 
event was in a metastable state, providing the right circumstances 
for Coleman--De~Luccia tunneling.)  Moreover, the evolution of the 
bubble universe after tunneling begins with $\dot{\phi} = 0$ and strong 
Hubble damping of $\dot{\phi}$ ($H \sim 1/t$), so at least in
simple models~\cite{Freivogel:2005vv,Dutta:2011fe} $\dot{\phi}^2$
is never large enough to interfere with the onset of slow-roll
inflation.

In this scenario, the universe would be homogeneous before the final 
stage of inflation.  Yet the universe immediately after tunneling 
would be an {\it empty}, curvature-dominated (open) universe.  To 
produce a matter-filled universe like the one in which we live, the 
tunneling would have to be followed by a period of slow-roll inflation. 
(Since the curvature term in the Friedmann equation falls off as $1/a^2$, 
after a time it can become dominated by vacuum energy, with $\rho_{\rm vac} 
\sim {\rm const}$.  Neither matter ($\rho_{\rm mat} \sim 1/a^3$) nor 
radiation ($\rho_{\rm rad} \sim 1/a^4$) can overtake the curvature 
term, except through an intermediate stage of vacuum energy domination.) 
Followed by the standard reheating process, the inflation would lead 
to a hot big-bang universe.  Moreover, if the duration of the slow-roll 
inflation were longer than $N \sim 60$ $e$-folds, then the flatness 
of the universe would be explained~\cite{Guth:1980zm,Linde:1981mu}, 
and the origin of structures could proceed as envisioned 
in Ref.~\cite{Mukhanov-Chibisov81}, and calculated in 
Ref.~\cite{Hawking:1982cz}.

The point we wish to emphasize is that inflation with what we consider 
a realistic form of $V (\phi)$, containing many local minima and hence 
many metastable states, would generically lead to multiple phases of 
inflation.  The observable properties of our universe today, as seen 
for example in the CMB, would be sensitive to the final phase of inflation, 
whereas details of the earlier processes would likely remain hidden from 
view, having been stretched far beyond the current horizon by the last 
$N \sim 60$ $e$-folds of inflation.  Given the well-known attractor 
behavior of slow-roll inflation~\cite{SalopekBond1990,Tegmark:2004qd}, 
quantitative predictions for observable quantities such as $\Omega_k$, 
$n_s$, $r$, and $\alpha = d n_s / d \ln k$ are essentially independent 
of anything that preceded the final phase of inflation.  Like any 
self-consistent effective field theory, inflation can be used to make 
specific predictions without knowing the exact behavior of the theory 
at arbitrarily high energies.  In particular, the predictions do not 
require knowledge of the prior phases of inflation~\cite{Tegmark:2004qd} 
or of Planck-scale physics~\cite{Cheung:2007st}.

ISL considered (in their Fig.~1d) the possibility that we discuss, 
with a tunneling episode prior to the slow-roll inflation, raising 
two criticisms with which we disagree.  First, they argue that this 
approach involves adding ``complicated features \ldots for the purpose 
of turning an unlikely model into a likely one.''  From our point of 
view, such ``complicated features'' are highly plausible in the context 
of the current understanding of particle theory.  ISL further argue 
that the plateau shape of the low-energy part of the potential is 
not a consequence of inflation, but instead is chosen only to fit the 
{\it Planck} data, a situation which they describe as ``trouble for 
the [inflationary] paradigm.''  It is of course true that inflation 
does not determine the shape of the potential, and indeed most 
inflationary theorists, including us, would consider a $\frac{1}{2} 
m^2 \phi^2$ or a $\lambda \phi^4$ potential to be a priori quite 
plausible for the low-energy part of the potential.  But this only 
means that (given current theoretical technology) the details of 
inflation will need to be determined by observation.  Many of the 
features of the standard model of particle physics are also determined 
by observation; this situation might suggest that some deeper theory 
underlies the standard model, but we do not think that it spells 
trouble for the standard model paradigm.

So far our arguments have depended only on the recognition that $V(\phi)$ 
might plausibly be a complicated function, with many local minima, as 
suggested by current ideas in particle theory, such as string theory. 
But once we consider a potential energy function with more than one 
metastable local minimum---or any potential energy function with a gentle 
plateau region---then eternal inflation seems unavoidable.  ISL refer 
to this as the ``multiverse problem.''  While we do not consider it a 
``problem,'' we agree that the multiverse is a very likely consequence. 
Regions filled with metastable ``vacua'' will generically inflate at 
a rate much faster than they decay, so the volume of inflating regions 
will grow exponentially as a function of the proper time, with no upper 
limit.  The metastable vacua will decay by bubble nucleation, producing 
``pocket universes'' at a rate that grows with the volume, and hence 
exponentially as a function of the proper time.

If this multiverse picture is combined with rather mild assumptions 
about anthropic selection effects, then it becomes very plausible 
that we live in a pocket universe which has undergone inflation, with 
no particular prejudice about whether the potential is plateau-like 
or not.  As described above, the pocket universe after tunneling 
would be a homogeneous open universe, with the scalar field that 
tunneled starting with $\dot{\phi} = 0$.  The amount of slow-roll 
inflation that follows depends on the shape of the potential.  Statistics 
alone would presumably favor small amounts of inflation if any, but 
Refs.~\cite{Freivogel:2005vv} and \cite{Guth:2012ww} argue that simple 
assumptions about the probability distribution for slow-roll potentials 
imply that the probability density for having $N$ $e$-folds of 
slow-roll inflation falls off for large $N$ only as a power of $N$: 
$P(N) \sim 1/N^p$, with $p \geq 0$ and $p \sim O(1)$.  Furthermore, 
Ref.~\cite{Freivogel:2005vv} argues there there is an anthropic minimum 
for the duration of the slow-roll inflation, $N \simgt 59.5$, based 
on the requirement that galaxy formation is possible.  We consider 
this a rather mild anthropic assumption, motivated by logic that 
parallels closely the logic of the anthropic bound on vacuum energy 
density~\cite{Weinberg:1987dv}---both vacuum energy and curvature 
suppress the growth of structure.

Although we have invoked an anthropic constraint on $N$, we 
emphasize that inflation is an essential part of the explanation. 
The anthropic constraint alone would give only $\Omega_k \simlt 
O(1)$.  To explain the observed flatness of the universe, 
$|\Omega_k| \simlt 0.01$~\cite{Ade:2013uln}, we need inflation, 
with its exponential sensitivity of $\Omega_k$ on $N$, $\Omega_k 
\propto e^{-2N}$.  Only about 2.5 $e$-folds of inflation beyond 
the minimum are needed to explain the observed level of flatness. 
According to the estimates of Refs.~\cite{Freivogel:2005vv} 
and~\cite{Guth:2012ww}, the relevant conditional probability 
is large; given that there is enough inflation to satisfy the 
anthropic cut ($N \simgt 59.5$), the probability that there 
is enough inflation to explain the observed level of flatness 
($N \simgt 62$) is very nearly one.

To summarize, the possibility that the final stage of inflation 
was preceded by a bubble nucleation event is at least one way that 
fine-tuning issues can be avoided.  The prior inflation in a metastable 
state can occur without any significant fine-tuning of initial 
conditions---all that is necessary is that the inflaton field 
roll down to a local minimum with positive energy density.  The 
tunneling must be followed by a period of slow-roll inflation, but 
with a complicated $V(\phi)$, as we would expect, it is very plausible 
that this occurs somewhere in field space.  Anthropic selection 
effects can then make it plausible that we live in a pocket universe 
that evolved in this way.  Moreover, one need not know how our 
observable universe came to undergo its final phase of inflation 
in order to make specific, quantitative predictions for observable 
quantities today.

\section{The Multiverse and Predictability}
\label{sec:multiverse}

ISL refer to a ``multiverse-unpredictability'' problem, and in the 
discussion they raise two issues.  First, they argue that the plateau 
potentials favored by {\it Planck} will lead to eternal inflation, 
and hence the measure problem~\cite{Guth:2000ka}.  We agree that if 
the observable inflation occurred on a plateau-like potential, eternal 
inflation seems very likely.  It can occur either while the scalar 
field is at or near the top of the plateau, or in a metastable state 
that preceded the final stage of inflation.  We also agree that this 
leads to the measure problem:\ in an infinite multiverse, we do not 
know how to define probabilities.  Since anything that can happen will 
happen an infinite number of times, the distinction between common 
events and extremely rare events requires a comparison of infinities, 
and that requires some method of regularization.  We do not yet know 
what is the correct method of regularization, or even what physical 
principles might determine the correct answer.  While we agree that 
this question is unanswered, we feel that acceptable measures (i.e., 
regularization prescriptions) have been proposed (e.g.~\cite{DeSimone:2008bq,%
Linde:1993xx,Bousso:2006ev,Nomura:2011dt,Garriga:2012bc}), and that the 
mere fact that we have not solved this problem is no reason to believe 
that nature would avoid eternal inflation.  Nature does not care 
whether we understand it or not.  However, since the measure problem 
is not fully solved, ISL are certainly justified in using their 
intuition to decide that eternal inflation seems unlikely to them. 
To us, the measure problem is simply an important problem that remains 
to be solved.

One reason for believing that the measure problem must be solved anyway 
is that the circumstances that lead to it are hard to avoid.  The 
cyclic model, for example, seems to also have a measure problem. 
In Ref.~\cite{Johnson:2011aa}, Johnson and Lehners study cyclic models 
that include a dark-energy-dominated phase, concluding that there is 
a measure problem, with probabilities that depend on a cut-off procedure 
for which there is no {\it a priori} way to determine.  They go on to 
claim that it is easier to find an acceptable measure in the context 
of cyclic cosmology, but the existence of the measure problem is not 
avoided.  While Johnson and Lehners confined their remarks to a subclass 
of cyclic models, we believe that the measure problem exists in all 
cyclic models.  The measure problem pertains to all probabilities, 
not just the probability of finding oneself in a particular phase. 
For any model in which anything that can happen will happen an infinite 
number of times, the measure problem applies.

The second issue that ISL include in the multiverse-unpredictability 
discussion is a claim that if there is a multiverse, then we should 
observe a large number of many-$\sigma$ deviations from predictions 
of our theories.  That is, they argue that inflation fails because 
it describes the data too well.  We emphasize that the existence of 
a measure problem does not mean that probability theory fails.  The 
different measures that have been proposed, and presumably the correct 
measure that we seek, obey all the standard properties of probability 
theory.  (Anything that can happen will happen, but {\it not} with 
equal probability.)  The predicted probabilities will depend on the 
measure, but we should expect that they will not differ radically 
from naive expectations, just as physicists in the 1920's could have 
expected that the emerging theory of quantum mechanics was not going 
to predict that cars should start tunneling out of their garages. 
That is, any acceptable extension of the laws of physics must be 
consistent with the older theories in the regime where the older 
theories have been tested successfully.  The measures discussed in 
Refs.~\cite{DeSimone:2008bq,Linde:1993xx,Bousso:2006ev,Nomura:2011dt,%
Garriga:2012bc} all fit this criterion.  Since all the basic axioms 
of probability theory are intact, an event with a probability of 
1/10 would be expected to occur about once in every 10 trials, 
as usual.

\section{Inflationary ``Unlikeliness Problem''}
\label{sec:unlikeliness}

ISL admit that inflationary plateau-like models obviously pass the test 
of giving predictions that agree with observation, thereby satisfying 
the criterion that is generally used to define the success of a theory. 
They argue, however, that this is not enough.  In what they call the 
inflationary ``unlikeliness problem,'' they contend that inflation 
occurring on a plateau is exponentially unlikely compared to inflation 
in a power-law potential.  As a simple example, they consider the 
plateau potential
\begin{equation}
  V(\phi) = \lambda (\phi^2-\phi_0^2)^2 \ ,
\label{plateau}
\end{equation}
which has a plateau for $|\phi| \ll \phi_0$, but behaves like a power-law 
potential for $|\phi| \gg \phi_0$.  They argue that there is a much 
larger range $\Delta \phi$ of scalar field values available in the 
power-law region, and a much larger maximum for the number $N_{\rm max}$ 
of $e$-folds of inflation, implying that inflation on the power-law part 
of the potential is exponentially more likely than inflation on the plateau.

In making this claim, ISL seem to have put themselves in the peculiar 
position of arguing, on the one hand, that eternal inflation leads 
to infinities, ``potentially rendering inflationary theory totally 
unpredictive,'' while at the same time arguing that they can tell us 
what inflation predicts, and that it is unambiguously at odds with 
the plateau behavior that the {\it Planck} observations favor.

At the level of inflation-scale physics, inflation on the power-law 
and the plateau parts of the potential of Eq.~(\ref{plateau}) are two 
distinct models, each perfectly consistent.  In comparing the likelihood 
of the two we need to consider Planck-scale physics, asking which 
inflation-scale scenario is more likely to develop from an assumed 
description at the Planck scale.  Since we and ISL agree that these 
models lead to eternal inflation, it is in this context that we will 
discuss their argument.  We believe that it is possible to make plausible 
predictions about Planck-scale issues in the context of eternal inflation, 
but they must be made carefully, choosing a probability measure which 
is at least free of known problems.  Unlike ISL, we would view the 
success or failure of such predictions not as a test of the inflationary 
paradigm, but rather as part of our exploration of the measure problem. 
So far predictions based on multiverse calculations have been pretty 
much limited to gross dynamical properties of the universe, such as 
the cosmological constant~\cite{DeSimone:2008bq,Bousso:2007kq} or 
$\Omega_k$~\cite{DeSimone:2009dq}.  Detailed particle physics issues, 
like the relative likelihood of finding a scalar field in one range 
of values vs.\ another, depend sensitively on the underlying particle 
dynamics, and do not appear to be even approachable at the present 
time.  Thus, if we assume that the final slow-roll inflation occurred 
in the potential of Eq.~(\ref{plateau}), in our view there is no way 
of knowing whether we should expect it to have occurred on the plateau 
or on the power-law part of the potential.

Nevertheless, ISL's argument for an ``inflationary unlikeliness problem'' 
sounds reasonable on a first reading, so we would like to look at it 
more carefully.  They argue that inflation on the power-law side of 
the potential is more likely because it allows a much larger range 
$\Delta \phi$ of the scalar field, and a much larger maximum number 
$N_{\rm max}$ of $e$-folds of inflation.  

We consider first the claim that the larger range $\Delta \phi$ of 
scalar field values implies that inflation on the power-law side is 
more likely.  If the physical system consisted of a single scalar 
field $\phi$ which started with random initial conditions at the 
Planck scale, then ISL's argument would be persuasive.  But the 
situation is not so clear if we consider the multiverse, or even 
if we just consider a more complicated system of scalar fields.%
\footnote{It is conceivable that string theory might not even allow 
 field values much larger than $M_{\rm Pl}$, in which case the large 
 range of $\Delta \phi$ on the power-law side would be an illusion.}
For example, if the plateau inflation was preceded by a tunneling 
event, then $\Delta \phi$ and $N_{\rm max}$ for the power-law side 
would have to be compared not with the corresponding values for 
the plateau, but instead with the corresponding values for the 
metastable state from which the tunneling occurred.  In this case 
ISL's argument for the larger amount of inflation on the power-law 
part of the potential completely disappears.

Furthermore, it is easy to construct models in which the plateau 
center, $\phi=0$, is an enhanced symmetry point, which can make 
it a likely endpoint of either tunneling or the stochastic evolution 
of scalar fields, even if the range of $\phi$ on the plateau is very 
small.  For example, with multiple scalar fields in the theory, it 
is quite plausible that there are metastable vacuum states for which 
the inflaton field $\phi=0$, with nonzero values for some or all 
of the other fields.  The dominant decay of such states could very 
plausibly maintain $\phi \approx 0$.  Similarly, if the potential 
energy function includes a term $\lambda \phi^2 \psi^2$, where 
$\psi$ is another scalar field which has a large value at early 
times, then $\phi$ can plausibly settle into its minimum energy 
state, $\phi \approx 0$, before $\psi$ becomes small.

Turning now to the claim that the probability of inflation on the 
power-law side is exponentially enhanced by the larger value of 
$N_{\rm max}$, we first point out that this argument also disappears 
if we assume that the final stage of inflation was preceded by a 
tunneling event.  But even if that is not the case, the issue of 
whether a large $N_{\rm max}$ leads to a large probability is precisely 
the kind of question that plays a major role in the discussion of 
measures, and hence must be handled with care.

The simplest measure, known as proper-time cutoff 
measure~\cite{Linde:1993nz,Linde:1993xx}, selects a finite sample 
spacetime volume of the multiverse by considering only events that 
occur before a final cutoff hypersurface that is chosen as a hypersurface 
of constant proper time.  The relative likelihood of events of different 
types is determined by counting the numbers of events in this sample 
spacetime volume, and then taking the limit as the final proper-time 
hypersurface is taken to infinity.%
\footnote{Proper time is of course not a globally defined quantity 
 in general relativity, so the meaning of a proper-time cutoff needs 
 to be described more carefully.  One begins by choosing an arbitrary 
 initial spacelike hypersurface of finite extent.  One then constructs 
 a congruence of geodesics that begin on the hypersurface and normal 
 to it, extending toward the future.  The sample spacetime region is 
 then chosen to be the region swept out by these geodesics, each followed 
 for a proper time $\tau = \tau_{\rm cutoff}$.  It is expected that 
 as $\tau_{\rm cutoff} \to \infty$, the resulting probabilities become 
 independent of the choice of the initial hypersurface.}

While the proper-time cutoff measure seems intuitive, it has been found 
to lead to a gross inconsistency with experience, often called the 
``youngness problem''~\cite{Linde:1994gy,Guth:2000ka-2,Tegmark:2004qd,%
Bousso:2007nd}.  The problem is driven by the huge disparity in time 
scales:\ the scale factors of the most rapidly inflating metastable 
false vacua are expected to have time constants of perhaps $\tau_{\rm min} 
\sim 10^{-37}~{\rm s}$, while the time scales relevant to the questions 
we ask might range from seconds to gigayears.  The growth of the sample 
spacetime volume is dominated by the most rapidly inflating vacua, so 
it is expected to grow as a function of $\tau_{\rm cutoff}$ with a time 
constant close to $\tau_{\rm min}$.  Since the growth is exponential, 
most of the spacetime volume will lie within a few time constants of 
the final hypersurface.  Thus, most of the pocket universes that form 
in the sample spacetime volume nucleate within a few time constants 
of the final hypersurface, and pocket universes that are older by 
some time interval $\Delta t$ are suppressed in probability by the 
smaller volume available at these earlier times, proportional to 
$e^{-3 \Delta t / \tau_{\rm min}}$.  Proper-time cutoff measure implies, 
therefore, that the statistical distribution of pocket universes is 
strongly biased toward very young universes.  Pocket universes as old 
as $\Delta t = 14$ billion years, for example, are suppressed by a 
factor such as $e^{-3 \Delta t / \tau_{\rm min}} \sim 10^{-{10^{55}}}$. 
Tegmark~\cite{Tegmark:2004qd} connects this strongly biased probability 
distribution to observation by estimating the probability that we 
find ourselves in a pocket universe old enough for the CMB temperature 
to be less than 3~K, finding that $P(T_{\rm CMB} < 3 \hbox{ K}) \sim
10^{-{10^{56}}}$.%
\footnote{Tegmark's result is more extreme than the probability of 
 $10^{-{10^{55}}}$ that we quote based on age, a discrepancy that appears 
 to be due to unimportant approximations.  The bottom line however is 
 clear:\ in proper-time cutoff measure, it is absurdly improbable for 
 us to find ourselves in a universe as old as what we see.}
Thus, the proper-time cutoff measure is emphatically ruled out by 
observation.

ISL recognize the failure of the proper-time cutoff measure, which 
they describe as ``weighting by volume,'' and yet their argument about 
a purported ``inflationary unlikeliness problem'' critically depends 
upon that flawed measure.  Measures that are currently considered 
acceptable, such as those in Refs.~\cite{DeSimone:2008bq,Linde:1993xx,%
Bousso:2006ev,Nomura:2011dt,Garriga:2012bc}, all avoid the youngness 
problem in one way or another, and they also satisfy several other 
consistency checks (see, for example, Ref.~\cite{Freivogel:2011eg}). 
All lead to the conclusion that the probability of finding oneself 
in a particular type of pocket universe is not enhanced by the amount 
of slow-roll inflation that occurs in that type of pocket universe.

Of the various known ways to avoid the youngness problem, perhaps the 
simplest to describe is the scale-factor time cutoff measure, which 
differs from the proper-time cutoff measure only by the choice of 
the time variable.  Instead of using proper time to define the final 
cutoff hypersurface, one uses scale factor time $t_{\rm sf}$, which 
in the context of a Friedmann-Robertson-Walker universe is just equal 
to the logarithm of the scale factor.  In an arbitrary spacetime, 
scale factor time can be defined as $1/3$ times the logarithm of 
the volume expansion factor.%
\footnote{In more detail, one begins by choosing an arbitrary initial 
 spacelike hypersurface of finite extent, as in proper-time cutoff 
 measure, and again one constructs a congruence of geodesics that begin 
 normal to this surface.  An interval of scale factor time along a 
 worldline is given by $d t_{\rm sf} = H d \tau$, where $d \tau$ is 
 the proper time interval and $H$ is the local expansion rate of the 
 geodesic congruence.  Equivalently, one could imagine filling the 
 initial hypersurface with a uniform density of dust particles that 
 subsequently follow the geodesics of the congruence.  Scale factor 
 time is then $-1/3$ times the logarithm of the density of dust.}
The volume of any comoving region then increases as $e^{3 t_{\rm sf}}$, 
so there is no youngness problem, since there is no large discrepancy 
in time scales, and regions with large values of $H$ are no longer 
given an enhanced weight.  Note that this measure is just as much 
volume-weighted as the proper-time cutoff measure---it is just the 
time coordinate that is different.  Once the cutoff is chosen, all 
volumes under the cutoff are counted equally.  Note also that we did 
not use the {\it Planck} data to influence our choice of measure; 
the absence of an enhancement from slow-roll inflation was implied by 
other---rather basic---considerations about probabilities, independent 
of any CMB data.

Without the strong exponential preference, the relative 
probabilities of the two starting points for the last stage of 
inflation---plateau-like or outer wall---become the issue of 
complicated dynamics in the multiverse, and we are unable to compute 
which will dominate with our current knowledge and technology.

To summarize, we have argued that there is no reason to conclude that 
inflation predicts that plateau inflation is unlikely.  We have indicated 
that the larger range $\Delta \phi$ of scalar field values allowed for 
power-law inflation is irrelevant if the final stage of plateau inflation 
is preceded by a tunneling event, and that in multifield models the 
plateau can be favored as an enhanced symmetry point.  We have also 
pointed out that in proposals for a probability measure that are 
currently considered acceptable, the larger amount of inflation that 
might be expected for power-law potential inflation does not lead to 
an enhanced probability.

\section{Inflation and the LHC}
\label{sec:LHC}

ISL close their paper with one final argument, which in this case is 
based on the LHC, rather than {\it Planck~2013}.  They argue that perhaps 
the absence of evidence for physics beyond the standard model should 
be extrapolated to a claim that there is no new physics up to the Planck 
scale.  In that case, given the measurements of the top quark and Higgs 
masses, the Higgs field potential energy function is predicted to reach 
a maximum, and then to decrease to a large negative vacuum energy density. 
While the barrier is high enough to give a lifetime for our vacuum that 
is large compared to the age since the big bang, ISL argue that it is 
highly unlikely for the Higgs field to end up in the tiny pocket around 
the correct electroweak breaking minimum, since there is a vastly larger 
region of field space in which it would roll toward Planck values.

As ISL noted themselves, the problem may evaporate if there is
new physics below the scale at which the Higgs potential energy
turns around.  For example, if there is supersymmetry around a
TeV scale or at a scale not too far from a TeV, e.g.\ a few
orders of magnitude above a TeV as in the scenario in
Ref.~\cite{Hall:2011jd}, then the problem is avoided. 
Alternatively, supersymmetry may exist at an intermediate
scale~\cite{Hall:2013eko}, which may also prevent the Higgs
potential from turning around.  (This is the case if the
supersymmetric threshold is slightly below the scale at which the
standard model Higgs quartic coupling would vanish.)  Nonminimal
couplings of the Higgs field, either to
gravity~\cite{Espinosa:2007qp} or to the
inflaton~\cite{Espinosa:2007qp,Lebedev:2012sy}, can also obviate
the problem.  It was also pointed out in
Ref.~\cite{Hertzberg:2012zc} that the turnaround of the Higgs
potential can be avoided by the Peccei-Quinn mechanism, which
with other assumptions can even lead to a phenomenologically
successful relationship between the Higgs mass and the dark
matter density.

Nevertheless, the suggestions of the previous paragraph are speculative, 
and it is possible that the Higgs potential actually does become negative 
above an intermediate scale, e.g.\ around $10^{11}~{\rm GeV}$, and 
that there is no new physics up to the Planck scale, as ISL suggest. 
We would argue that, even in this case, there is no problem for 
inflation in the context of an eternally inflating multiverse.  The 
key issue is that there is no plausible way that regions in which 
the Higgs field has run off to Planckian values could support life. 
The large negative vacuum energy density is enough to ensure that 
these regions would collapse to a crunch on time scales far shorter 
than a second, leaving only those (initially very rare) regions where 
the Higgs field has rolled toward small values.  It has always been 
assumed that the multiverse includes a large number of types of 
pocket universes that do not support life, so the possibility 
described by ISL merely adds one to that number.  For the multiverse 
framework to be consistent, it is only necessary that the probability 
that intelligent observers find themselves in a pocket universe like 
ours is not unreasonably small.

For the Higgs field to remain in the region within the potential
maxima during inflation, there are constraints on various
inflationary parameters, derived in references cited by
ISL~\cite{Espinosa:2007qp,Kobakhidze:2013tn}, that must be
obeyed.  In particular there are constraints on the energy scale
of inflation, on the amplitude of tensor fluctuations, and on the
amplitude of density perturbations for the case of power-law
potentials, but none of these pose trouble for inflationary
models.

While we see no reason to be concerned with the case described by 
ISL---the case in which the standard model holds exactly up to the 
Planck scale, with a Higgs potential that turns negative---in the 
context of the multiverse there are other interesting possibilities. 
One could imagine, for example, a vacuum in the landscape for which 
the physics is given by the standard model, except for an offset 
in the vacuum energy density which makes the value very near zero 
when the Higgs field is at the Planck scale.  Such a universe would 
still be inhospitable to life:\ with the values of gauge and Yukawa 
couplings taken from the standard model, a Planck-scale Higgs 
expectation value would make all the standard model particles 
so heavy that they would presumably not even be created during 
reheating, leaving a universe populated only by photons (and 
possibly neutrinos).

One might argue that if the Yukawa couplings vary, becoming
vanishingly small in such a way that the masses of the quarks and
leptons are fixed at their standard model values, then the
resulting universe might not be very much different from
ours~\cite{Harnik:2006vj}.  In this case, the probability of
finding ourselves in such a universe is limited by the
probability of obtaining such tiny couplings in the landscape,
which is likely to be small. In addition, there may be anthropic
reasons associated with the absence of weak interactions that
prevent life in such a universe, despite the analysis of
Ref.~\cite{Harnik:2006vj}.  In any event, while one may continue
to speculate about conceivable vacua, the exercise would only
pose trouble for the inflationary paradigm if someone identified
a class of vacua in the landscape which could be shown to
strongly dominate over our vacuum in probability.  This has not
happened.

\section{Conclusions}
\label{sec:concl}

Inflationary cosmology rests on very firm foundations.  Rather than relying 
on untested (though certainly interesting) speculations about additional 
spatial dimensions or repeated collisions of hypothetical branes, inflation 
builds upon decades of in-depth study of quantum field theory in curved 
spacetime.  Like many other successful modern physical theories, inflation 
may be understood as an effective field theory, capable of making specific, 
quantitative predictions for observables in various energy regimes of 
interest, even in the absence of complete knowledge of physics at arbitrarily 
high energy scales.  Many of those quantitative predictions have been 
subjected to empirical tests across a wide range of experiments, including 
most recently with the {\it Planck} satellite.  Every single test to date 
has shown remarkably close agreement with inflationary predictions.

We agree with Ijjas, Steinhardt, and Loeb~\cite{Ijjas:2013vea} that 
important questions remain.  A well-tested theory of physics at the Planck 
scale remains elusive, as does a full understanding of the primordial 
singularity and of the conditions that preceded the final phase of 
inflation within our observable universe.  Likewise, although significant 
progress has been made in recent years, a persuasive theory of 
probabilities in the multiverse has not yet been found.  We strongly 
disagree with ISL, however, that these remaining challenges represent 
any sort of shortcoming of inflationary cosmology.  Quite the opposite:\ 
the inflationary paradigm, with its many successes, provides a framework 
within which such additional questions may be pursued.

In assessing the criticisms of inflation by ISL, we have identified 
several assumptions in their arguments that we consider problematic. 
Most stem from an outdated view in which a single phase of inflation 
is assumed (or required) to persist from the Planck scale to the 
inflationary scale.  None of the quantitative predictions from 
inflationary cosmology for various observables require such an 
assumption, nor does such an assumption seem at all realistic in 
the light of recent developments in high-energy theory.

Recent experimental evidence, including the impressive measurements with 
the {\it Planck} satellite of the CMB temperature perturbation spectrum 
and the strong indication from the LHC that fundamental scalar fields such 
as the Higgs boson really exist, put inflationary cosmology on a stronger 
footing than ever.  Inflation provides a self-consistent framework with 
which we may explain several empirical features of our observed universe 
to very good precision, while continuing to pursue long-standing questions 
about the dynamics and evolution of our universe at energy scales that 
have, to date, eluded direct observation.

\begin{acknowledgments}
We would like to thank Bruce Bassett, Andrei Linde, and Alex
Vilenkin for helpful conversations.  We understand that Linde is
preparing a manuscript on his 2013 Les Houches
lectures~\cite{Linde-Houches}, which also contains
counterarguments to ISL in the appendix; see also his 2013 lecture at the
KITP~\cite{Linde-KITP}. This work was supported in part by the
U.S.\ Department of Energy (DOE) under cooperative research
agreement DE-FG02-05ER41360.  The work of Y.N. was also supported
in part by the Director, Office of Science, Office of High Energy
and Nuclear Physics, of the US Department of Energy under
Contracts DE-AC02-05CH11231, the National Science Foundation
under grant PHY-1214644, and the Simons Foundation grant 230224.
\end{acknowledgments}

\end{document}